\documentclass{article}
\usepackage{spconf,amsmath,graphicx}
\usepackage{booktabs}
\usepackage{multirow}

\title{FRAME-LEVEL EMOTIONAL STATE ALIGNMENT METHOD FOR SPEECH EMOTION RECOGNITION}
\name{Qifei Li, Yingming Gao, Cong Wang, Yayue Deng, Jinlong Xue, Yichen Han, Ya Li}
\address{Beijing University of Posts and Telecommunications, Beijing, China}
\begin{document}
%
\maketitle
\begin{abstract}
Speech emotion recognition (SER) systems aim to recognize human emotional state during human-computer interaction. Most existing SER systems are trained based on utterance-level labels. However, not all frames in an audio have affective states consistent with utterance-level label, which makes it difficult for the model to distinguish the true emotion of the audio and perform poorly. To address this problem, we propose a frame-level emotional state alignment method for SER. First, we fine-tune HuBERT model to obtain a SER system with task-adaptive pretraining (TAPT) method, and extract embeddings from its transformer layers to form frame-level pseudo-emotion labels with clustering. Then, the pseudo labels are used to pretrain HuBERT. Hence, the each frame output of HuBERT has corresponding emotional information. Finally, we fine-tune the above pretrained HuBERT for SER by adding an attention layer on the top of it, which can focus only on those frames that are emotionally more consistent with utterance-level label. The experimental results performed on IEMOCAP indicate that our proposed method performs better than state-of-the-art (SOTA) methods. The codes are available at github repository\footnote{https://github.com/ASolitaryMan/HFLEA.git}.
\end{abstract}
\section{Introduction}
\label{sec:intro}

In order to improve the experience of human-computer interaction, speech emotion recognition has become one of the research hotspots in recent years. Technologies in this field have advanced considerably over the past decade.

The conventional methods for SER focus on using neural networks to mine emotional information from hand-crafted or spectral features~\cite{mirsamadi2017automatic, li2019dilated, rajamani2021novel, wu2022neural}. 
Due to limited labeled data, these methods have only shown slight performance improvements. With the success of natural language processing pretraining models~\cite{kenton2019bert}, several self-supervised audio pretraining models have emerged, such as wav2vec~\cite{schneider19_interspeech}, wav2vec2.0~\cite{baevski2020wav2vec}, HuBERT~\cite{hsu2021hubert}, and WavLm~\cite{chen2022wavlm}. Those models are obtained by self-supervised pretraining using large amounts of unlabelled data, and what they learn can be transferred to improve the performance of downstream tasks, such as automatic speech recognition~\cite{hsu2021}, SER~\cite{shen2023mingling}, speaker recognition~\cite{gat2022speaker}, etc.

There are mainly three methods for implementing SER using audio pretrained models. The first is to extract the embeddings of the pretrained model as the input of the downstream task model~\cite{shen2023mingling, kakouros2023speech}. The second kind is to fine-tune the model for SER~\cite{pepino21_interspeech,chen2023exploring,xia2021temporal}. The third category involves redesigning the pretext task, pretraining the model based on this pretext task, and then fine-tuning it to implement SER~\cite{chen2023exploring}. Based on the second and third methods, Chen et al.~\cite{chen2023exploring} utilize wav2vec2.0 to realize a pseudo label task adaptive pretraining approach (P-TAPT) for sentiment analysis, which can align frame-level pseudo-emotion labels with frames to alleviate the inconsistency between emotional states of some frames in the audio and its utterance-level label, and performs better than many SER systems that only utilize utterance-level labels. 

However, P-TAPT has two aspects that can be improved. The first point is that it aims to achieve a frame-level emotional alignment method similar to that of HuBERT. However, it only introduces frame-level pseudo-emotion labels and fine-tunes wav2vec2.0 directly to realize frame-level emotion state alignment instead of pretraining HuBERT. The HuBERT has been proven to be more suitable for this offline discrete frame-level self-supervised task~\cite{hsu2021hubert, qian2022contentvec}. The second one is that the authors fine-tuned the aligned model directly with average pooling to aggregate frame-level representations to utterance-level representations for final utterance-level SER system. This method may not fully exploit the aligned frame-level emotional information.

\begin{figure*}[ht]
    \centering
    \includegraphics[scale=0.8]{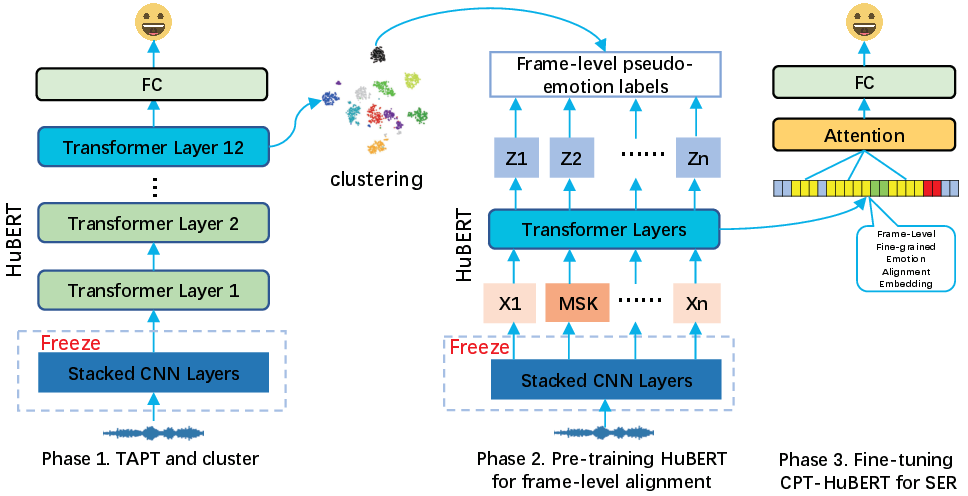}
    \caption{Framework of the proposed FLEA. MSK and FC denote the mask and the fully connected layer respectively.}
    \label{fig:fig1}
\end{figure*}

Inspired by~\cite{hsu2021hubert, chen2023exploring, qian2022contentvec}, we propose a frame-level emotion alignment (FLEA) method for SER, which is an extension and improvement of P-TAPT. In the first step, we fine-tune the HuBERT model to implement a SER system and extract the embeddings of its i-$th$ transformer layer for clustering to achieve frame-level pseudo-emotion labels. Then, we continue to pretrain the HuBERT model using pseudo-labels, referred to as CPT-HuBERT that allows to automatically align frames with pseudo-emotion labels. Finally, we add an attention layer on the top of CPT-HuBERT and fine-tune it for SER. The role of attention is to align frame-level representations with utterance-level labels. In addition, we explore the effect of using different transformers layers of embeddings to cluster and different number of clusters on SER performance. We perform experiments on IEMOCAP~\cite{busso2008iemocap} to validate the effectiveness of FLEA. The unweighted accuracy (UA) and weighted accuracy (WA) of FLEA are 75.7\% and 74.7\%, outperforming SOTA methods.

\section{PROPOSED METHOD}
\label{sec:method}
In this section, we first review HuBERT model, which is the key backbone of our proposed method. Then, we introduce our proposed method in detail, and the system framework is shown in Fig.\ref{fig:fig1}.

\subsection{The Review of HuBERT}
\label{hubert}
HuBERT is a large pretrained model obtained through self-supervised learning, which is used to learn general audio representations from unlabeled raw audio signals for various downstream speech tasks. First, HuBERT leverages the offline clustering algorithm to generate pesudo labels for mask language model (MLM) pretraining. Second, the raw audio signals need to be encoded into meaningful continuous latent representations by a feature extractor which consist of a stack of CNN layers. Finally, the model utilizes the transformer layers to learn the structure of spoken inputs by predicting the cluster for masked audio segments. The training of HuBERT includes two phases. The pseudo labels used in the first phase of pretraining are generated from the mel frequency cepstrum coefficient (MFCC), and the pseudo labels used in the second phase of pretraining are generated from the embeddings of the model saved in the first phase. The MLM pretraining means that the representations of the masked frames and unmasked frames require computing cross-entropy losses with pseudo-labels respectively. The two losses denote as $L_m$ and $L_u$, and the final predictive loss $L$ of the model can be calculated as follows:
\begin{align}
  L = \alpha{L_m} + (1-\alpha){L_u}
  \label{equation:eq1}
\end{align}
The predictive loss forces the model to learn good high-level representations of unmasked frames to help infer the pseudo labels of masked frames correctly~\cite{kenton2019bert, hsu2021hubert}.

\subsection{TAPT and Pretraining HuBERT}
\label{sec:pretrain and cluster}

\textbf{TAPT and cluster.} Since all of the SER datasets only have utterance-level labels, in order to achieve frame-level affective alignment, we need to introduce frame-level pseudo-emotion labels. To this end, we generate pseudo labels by following this methods~\cite{chen2023exploring,xia2021temporal} which prove that frame-level emotion state can be inferred by training with a segment-based classification objective. As shown in phase 1 in Fig.~\ref{fig:fig1}, we first fine-tune HuBERT for SER with TAPT~\cite{gururangan2020don}. TAPT is kind of fine-tune, which first continues pretraining pretrained model on the target datasets to bridge the gap between pretrained and target domains and then fine-tunes above pretrained model for SER. Next, we extract the embeddings from the i-$th$ transformer layer in HuBERT to generate frame-level pseudo-emotion labels by k-means algorithm. Consistency in the k-means mapping from audio input to discrete targets is crucial, as it allows the model to focus on modeling the emotion sequential structure of the inputs during MLM pretraining.

\textbf{Pretraining HuBERT.} According to the theory that bad teachers (pseudo labels) make good students (learned representations)~\cite{hsu2021,chen2023exploring,qian2022contentvec}, we continue using MLM to pretrain the HuBERT with the frame-level pseudo-emotion labels as illustrated in Fig.~\ref{fig:fig1}. During pretraining, we set the value of $\alpha$ in the predictive loss to be 1 as in the official HuBERT. In other words, we only apply the loss function over the masked regions. After pretraining, the embeddings of each frame in the last transformer layer of CPT-HuBERT is mapped to a discrete frame-level pseudo-emotion labels to achieve frame-level fine-grained emotion alignment embeddings. 

\subsection{Soft Attention}
\label{sec:attention}
An advantage of the frame-level fine-grained emotion alignment embeddings is that there are clear differences between the frame embeddings of different emotions, while the embeddings of frames representing the same emotion are typically adjacent to each other, as shown in the third phase of Fig.\ref{fig:fig1}. By leveraging a simple attention, we can focus on frames that are strongly related to the utterance-level emotion label, while disregarding frames that are irrelevant to the emotion label. This approach effectively addresses the issue of interference from emotion label-unrelated frames in SER.

In this work, we use the soft attention to align the frame-level fine-grained emotion alignment embeddings with utterance-level emotion labels. The attention is implemented as follows:
\begin{align}
  \alpha_i = softmax(tanh(\mathbf{W}x_i))
  \label{equation:eq2}
\end{align}
\begin{align}
  Z = \sum^{N}_{i=1}\alpha_ix_i
  \label{equation:eq3}
\end{align}
where the $x_i$ is the frame-level fine-grained emotion alignment embedding of the i-$th$ frame and the $\mathbf{W} \in R^{1 \times D}$ is a trainable parameter to encode the attention weights of frames. The variables $\alpha_i$, $N$, $Z$ and $D$ are the attention weights of the i-$th$ frame, the number of frames, the utterance-level emotional representation and the dimension of embeddings respectively.

\section{EXPERIMENTS}
\label{sec:exp}

\subsection{Experimental Setup}
\label{sec:setup}
\textbf{Dataset.} IEMOCAP~\cite{busso2008iemocap} is a well-known multi-modal emotion corpus, which includes audio, visual and lexical modalities. In this work, we only use the data of audio modality. The corpus contains five recording sessions, each session has one male and one female speaker. In order to prevent leakage of speaker information and labels, whether it is pre-training HuBERT or fine-tuning the model for SER, we perform five-fold cross-validation using a leave-one-session out strategy on the corpus. This means that the data of four sessions are used as training data; the data of one speaker from the remaining session is used as the validation set and the data of other one speaker forms the testing set. We conduct our experiments with the 5531 audio utterances of four emotion categories \textit{happy (happy \& excited), sad, neutral} and \textit{angry}.

The UA and WA are used as evaluation metrics in line with previous methods. The UA is the mean of the accuracy of each category and WA is the accuracy of all samples.

\textbf{Experimental Details.} The pretrained HuBERT\footnote{https://huggingface.co/facebook/hubert-base-ls960} we used is the backbone of FLEA, which consists of 6 CNN layers and 12 transformer layers. It has an embedding dimension of 768. When we fine-tune HuBERT for SER with TAPT, we use the official k-means model\footnote{https://github.com/facebookresearch/fairseq/tree/main/examples/hubert} to generate pseduo labels to continue pretraining HuBERT on the IEMOCAP. In the first and third phases as shown in Fig.\ref{fig:fig1}, whether TAPT or fine-tuning the model for SER, the batch size is 64, the learn rate is 1$e$-4, the loss function is cross entropy loss and the optimizer is AdamW. The batch size is 64 and the epochs are all 40. In order to explore the impact of clustering on the final SER performance, we follow the official HuBERT~\cite{hsu2021hubert} extracted the embeddings of the 6-$th$, 9-$th$, and extra 11-$th$ transformer layers of HuBERT and performed clustering with 50, 100, and 150 clusters for each layer of embeddings. The reason we chose these numbers of clusters is the small sample size of the dataset.

During pretraining HuBERT, the mini-batches are generated by bucket sampler. The learning rate is 5$e$-4, the training steps are 20,000, and the warmed up steps are 4,000. More details are shown in our github project. The P-TAPT is used as baseline~\cite{chen2023exploring}.

\begin{table}[pt]
    \centering
    \caption{The UA/WA (\%) of ablation experiments of different poolings and different clusters of different transformer layers. BL means Baseline. The i-$th$ layer represents extracting the embeddings from the i-$th$ transformer layer to cluster.}
    \label{tab:tab1}
    \begin{tabular}{@{}cccc@{}}
    \toprule
    Layers              & Clusters & Average Pooling & Attention Pooling \\ \midrule
    BL~\cite{chen2023exploring}                  & -        & 74.3/-       & -                 \\ \midrule
    TAPT                & -        & 74.1/72.8       & -                 \\ \midrule
    \multirow{3}{*}{6-$th$}& 50       & 75.0/73.6       & 75.2/73.6        \\
                        & 100      & 74.8/73.3       & 75.1/73.5         \\
                        & 150      & 74.5/72.7       & 74.3/73.2         \\ \midrule
    \multirow{3}{*}{9-$th$}& 50       & \textbf{75.1}/73.5       & \textbf{75.7}/\textbf{74.7}         \\
                        & 100      & 75.0/\textbf{73.9}       & 75.3/74.0         \\
                        & 150      & 74.8/73.5       & 74.6/73.2         \\ \midrule
    \multirow{3}{*}{11-$th$}& 50       & 74.3/72.7       & 74.4/73.0         \\
                        & 100      & 74.0/72.8       & 74.2/72.7         \\
                        & 150      & 74.3/70.1       & 73.5/72.5         \\ \bottomrule
    \end{tabular}
\end{table}

\subsection{Experiments and Analysis}
\label{sec:ablation}
A series of ablation experiments are conducted to evaluate the effect of pretraining HuBERT for frame-level emotion state alignment, attention and the number of clusters produced by different layers of embeddings on the performance of SER system. 
For comparison with the baseline, we list the results of FLEA with attention pooling, and the results of the same fine-tuning method as baseline, which fine-tunes HuBERT for SER with using average pooling to aggregate embeddings, as shown in Table~\ref{tab:tab1}. 

\subsubsection{The impact of embeddings and clusters}
From Table~\ref{tab:tab1}, we can observe that different layers of embeddings and number of clusters have different effects on SER performance. No matter how many classes are clustered, the effect of pretraining HuBERT for SER with the frame-level pseudo-emotion labels clustered by the embedding of the 9-$th$ layer is better than that of the 6-$th$ and 11-$th$ layers. This experimental results demonstrate that the embedding of the 9-$th$ layer is more suitable for generating pseudo labels to pretrain HuBERT. The finding is consistent with the study~\cite{hsu2021hubert}. 

In addition, regardless of which layer of embeddings are used to cluster, the performance of SER decreases as the number of clusters increase. This phenomenon may be related to the size of the dataset. A clustering number of 50 is appropriate on the IEMOCAP dataset for pretraining HuBERT. However, we believe this is due to the limitation of dataset size. With more sentiment data, this cluster number may be larger and the model would be more robust, as in the case of the official HuBERT clustering number of 500.

\subsubsection{The role of pretraining HuBERT and attention pooling}
As shown in Table~\ref{tab:tab1}, in the 9-$th$ layer, fine-tuning HuBERT with average pooling for SER performs better than the baseline regardless of the numbers of clusters. This indicates that using the MLM method to pretrain HuBERT to realize frame-level emotion state alignment yields better performance than directly fine-tuning wav2vec2.0 for the same purpose.

Furthermore, at a cluster number of 50, the UA and WA of FLEA are improved by 0.8\% and 1.6\%, respectively, compared to those of fine-tuned CPT- HuBERT with average pooling for SER, and by 1.9\% over the UA of the baseline. This indicates that introduction of attention to align frame-level embeddings with utterance-labels is effective. Compared to average pooling, attention pooling can pay attention to the those frames strongly related to utterance-label, which makes better use of aligned frame-level emotion information. Although the performance of average pooling gradually approaches or even exceeds that of attention pooling as the number of clusters increases, the performance of the final SER also decreases.

\subsection{Performance Comparison with previous Methods}

The effectiveness of our proposed method can be spotlighted via comparing with current key results performed on the IEMOCAP corpus (Table~\ref{tab:tab2}). It shows that the best UA (75.7\%) and WA (74.7\%) are achieved by our proposed method. Moreover, our system outperforms the baseline\footnote{https://github.com/b04901014/FT-w2v2-ser} even though it leaks speaker information while fine-tuning wav2vec for clustering to generate frame-level pseudo labels. After our testing, FLEA performs better if we use the baseline clustering approach, which leaks speaker information. In addition, as shown in Table~\ref{tab:tab2}, our method performs well beyond the latest SER methods of the day, such as ShiftCNN, SUPERB, SMW-CAT, etc. Meanwhile, the performance of FLEA is close to some multi-modal methods, which are based on the modalities of audio and lexical.

\section{CONCLUSIONS}
\label{sec:conclusion}
In this work, we propose a novel method called FLEA for SER, which achieves SOTA performance on the IEMOCAP corpus. We show that frame-level emotion state alignment can be achieved by pre-training HuBERT with MLM method using frame-level pseudo-emotion labels. On the above aligned model, performing attention pooling to aggregate frame-level embeddings to utterance-level embeddings can get better performance for SER. Furthermore, we find that the model pre-trained for SER using the pseudo label generated by the embedding clustering of the 9-$th$ transformer layer of HuBERT has the best performance and the most robustness. A cluster number of 50 is best suitable on IEMOCAP, but it may not be robust for other corpora due to the size of dataset. In future work, we will explore the relationship between dataset size and number of clusters.

\begin{table}[pt]
\caption{Performance comparison of UA and WA with previous methods on IEMOCAP. The P-TAPT is baseline.}
\label{tab:tab2}
\begin{tabular}{@{}ccccc@{}}
\toprule
Type                   & Year & Methods   & UA(\%)        & WA(\%)        \\ \midrule
\multirow{5}{*}{Audio} & 2023 & P-TAPT~\cite{chen2023exploring} & 74.3          & -             \\
                       & 2023 & SMW-CAT~\cite{he2023multiple}  & 74.2          & 73.8          \\
                       & 2023 & ShiftCNN~\cite{shen2023mingling} & 74.8          & 72.8          \\
                       & 2023 & SUPERB~\cite{kakouros2023speech}   & 75.6          & -             \\
                       & -    & \textbf{Ours}     & \textbf{75.7} & \textbf{74.7} \\ \midrule
\multirow{2}{*}{Multi-modal} & 2023 & MTG~\cite{he2023xx}  & 75.0          & 74.5          \\
            & 2023 & MSMSER~\cite{wang2023exploring}   & 76.4          & 75.2          \\ \bottomrule
\end{tabular}
\end{table}

\section{ACKNOWLEDGEMENTS}
\label{sec:acknowledgement}
The work was supported by the National Natural Science Foundation of China (No.\ 62271083), the Special Fund for Military Healthcare Committee (No.\ 22BJZ28), the Fundamental Research Funds for the Central Universities (No.\ 2023RC13), open research fund of The State Key Laboratory of Multimodal Artificial Intelligence Systems (No.\ 202200042, No.\  202200012) and BUPT Excellent Ph.D. Students Foundation (No.\ 2023116)

\vfill\pagebreak



\small
\bibliographystyle{IEEEbib}
\bibliography{strings,refs}

\end{document}